\documentclass[aps,pre,twocolumn,showpacs,showkey,square,numbers,amssymb,amsmath,nobibnotes]{revtex4-1}
\usepackage{bm}
\usepackage{times}
\usepackage{graphicx}
\usepackage[usenames,dvipsnames,svgnames,table]{xcolor}
\usepackage{hyperref,float,subfigure}
\hypersetup{colorlinks=true, linkcolor=NavyBlue, citecolor=PineGreen,urlcolor=cyan}
\usepackage{mathtools}

\newcommand{\bM     }{\mbox{\boldmath$M$}}

\newcommand{\bxi     }{\mbox{\boldmath$\xi$}}

\newcommand{\bracket}[1]{\left\langle #1\right\rangle}
\newcommand{\beeq}[1] {\begin{equation}\begin{split}#1\end{split}\end{equation}}

\begin{document}
\title{Large deviation theory for diluted Wishart random matrices}
\author{Isaac P\'erez Castillo}
\address{Department of Quantum Physics and Photonics, Institute of Physics, UNAM, P.O. Box 20-364, 01000 Mexico City, Mexico}
\address{London Mathematical Laboratory, 14 Buckingham Street, London WC2N 6DF, United Kingdom}
\author{Fernando L. Metz}
\address{Institute of Physics, Federal University of Rio Grande do Sul, 91501-970 Porto Alegre, Brazil}
\address{Physics Department, Federal University of Santa Maria, 97105-900 Santa Maria, Brazil}
\address{London Mathematical Laboratory, 14 Buckingham Street, London WC2N 6DF, United Kingdom}

\begin{abstract}
Wishart random matrices with a sparse or diluted structure are ubiquitous in the processing of large datasets, with applications in physics, biology and economy. In this work we develop a theory for the eigenvalue fluctuations of diluted Wishart random matrices, based on the replica approach of disordered systems. We derive an analytical expression for the cumulant generating function of the number of eigenvalues  $\mathcal{I}_N(x)$ smaller than  $x\in\mathbb{R}^{+}$, from which all cumulants of $\mathcal{I}_N(x)$ and the rate function $\Psi_{x}(k)$ controlling its large deviation probability $\text{Prob}[\mathcal{I}_N(x)=kN]  \asymp e^{-N\Psi_{x}(k)}$ follow. Explicit results for the mean value and the variance of $\mathcal{I}_N(x)$, its rate function, and its third cumulant are discussed and thoroughly compared to numerical diagonalization, showing a very good agreement. The present work establishes the theoretical framework put forward in a recent letter [Phys. Rev. Lett. {\bf 117}, 104101] as an exact and compelling approach to deal with eigenvalue fluctuations of sparse random matrices.
\end{abstract}
\maketitle


\section{Introduction}

In the last decades we have experienced an explosion of available information - the so-called Big Data era. Problems in modern data analysis usually involve a large number of variables and observations, posing new challenges in the processing of data. This high-dimensionality of the dataset typically occurs in climate studies, genetics, biomedical imaging, and economics \cite{Fan2014}.

Suppose one performs $P$ measurements of $N$ variables characterizing a system. For instance, the variables could be assets in a stock market or a collection of climate observables, while the measurements of all variables could be simultaneously performed for $P$ different times. The collected data can be organized in an $N \times P$ matrix $\bxi$, with the element $\xi_{ij}$ providing the measurement $j$ of the variable $i$. The $N \times N$ sample covariance matrix $\bM$ of the dataset is built from the product $\bM = \bxi \bxi^T$ and it encodes all possible correlations among the variables. The covariance matrix is at the core of multivariate statistical analysis, with applications in dimensional reduction methods and classifying procedures, such as principal component \cite{bookPCA} and linear discriminant analyses \cite{bookLDA}, respectively.

Generally speaking, it is reasonable to expect that in many natural phenomena each variable is significantly correlated with only a few others, giving rise to sparse covariance matrices, whose main feature is the presence of a large amount of entries that are very small or even zero. In this context, an important example is the problem of inferring, from the empirical covariance matrix, the causal influences among the individual components of a system. This is typically the case, for instance,  in the experimental reconstruction of the interactions between the elements of biological systems, such as cellular signaling networks \cite{Sachs2005}, gene regulatory networks \cite{Butte2000} and ecological association networks \cite{Deng2012,Kurtz2015}. Besides that, performing numerical tasks with large covariance matrices, where all entries are strictly nonzero, is computationally very expensive and, in this case, one usually resorts to regularization techniques in order to bring the matrix into a sparse form \cite{Fanreview}.

Since the pioneering work of John Wishart  \cite{Wishart}, random  matrix theory has been playing a pivotal role in multivariate statistics \cite{Guptabook}.  Essentially, results derived from random matrix models serve as important benchmarks through which comparisons with real data can be made. The simplest null model for the covariance matrix $\bM$ consists in assuming that the entries of $\bxi$ are independent Gaussian random variables. In this case, the joint distribution of the eigenvalues of $\bM$ is analytically known, which forms the starting point to employ the Coulomb gas technique \cite{Dyson1,Dyson2,Dyson3} and derive a wealth of quantitative information about the eigenvalue statistics of the covariance random matrix \cite{Vivo2007,Isaac2010,Vivo,Majumdar2009}, including a detailed account of the typical and atypical fluctuations of its eigenvalues \cite{Vivo}. Unfortunately, apart from the averaged spectral density \cite{Perez2008}, much less is known about the eigenvalue statistics of sparse covariance matrices. The main reason lies in the absence of an analytical expression for the joint distribution of eigenvalues, which hampers the application of the Coulomb gas approach. This is a general problem in ensembles of diluted random matrices and, although novel approaches have helped us to push forward the understanding of the eigenvalue statistical properties in these ensembles \cite{Perez2008,Kuhn2008,Perez2009,Perez2010,MetzA2010,MetzA2011,MetzA2012,MetzA2013,MetzA2014,MetzA2016}, from an analytical viewpoint, we still have a long way to go when compared to classical ensembles of random matrices.

In this work we develop an analytical approach to quantify the eigenvalue fluctuations of sparse covariance random matrices. By relying on a recent technique  \footnote{These techniques were developed in \cite{Metz2015,Metz2016,Metz2017}. See also \cite{Coolen2016} for a similar technique introduced in a different context.}, based on the replica method of disordered systems \cite{BookParisi}, we derive an analytical expression for the large-$N$ behavior of the cumulant generating function of the number of eigenvalues $\mathcal{I}_N(x)$ smaller than a certain threshold  $x\in\mathbb{R}^{+}$. This function gives access to all cumulants of the random variable $\mathcal{I}_N(x)$ as well as to its large deviation probability, providing a full picture of the eigenvalue fluctuations for this class of random matrices. From the numerical solution of our analytical equations, we present explicit results for the mean and the variance of $\mathcal{I}_N(x)$, its third cumulant, and its rate function governing the large deviation probability. We show that, similarly to the atypical eigenvalue fluctuations of the adjacency matrix of random graphs \cite{Metz2016}, the rate function of $\mathcal{I}_N(x)$ is asymmetric around its minimum, which characterizes an unbalanced contribution of rare samples responsible for increasing or decreasing $\mathcal{I}_N(x)$ with respect to its typical value. The exactness of our theoretical findings is fully supported by direct diagonalization of finite random matrices.

In the next section we define the ensemble of sparse Wishart matrices and the main quantity we consider in order to study eigenvalue fluctuations. In section \ref{sectionresults} we present explicit results derived from our theoretical approach, while in the last section we discuss some final remarks. All technical details are concentrated on two appendices. Appendix \ref{AppA} explains all steps involved in the derivation of the cumulant generating function of $\mathcal{I}_N(x)$. In this appendix we also discuss some mathematical subtleties regarding the representation of $\mathcal{I}_N(x)$ in terms of the complex logarithm. In appendix \ref{Weighted} we describe the algorithmic approach employed to solve numerically the main analytical equations obtained from the theory.


\section{Random matrix model and the general setting}

As we are interested in the ensemble of sparse or diluted Wishart matrices, we need to decide a way to introduce dilution in the Gaussian or classical Wishart ensemble.  We proceed as follows: we consider rectangular $N\times P$ matrices $\bm{\xi}$ whose entries are independent and identically distributed random variables drawn from the distribution
\beeq{
  p(\xi_{i\mu})=\frac{d}{N}P_{\xi}(\xi_{i\mu})+\left(1-\frac{d}{N}\right)\delta(\xi_{i\mu})\,,
  \label{aa1}
}
where $P_{\xi}(\xi)$ is the probability density for the nonzero entries of $\bm{\xi}$. From a graph viewpoint \cite{Bollobas}, the random
matrix $\bm{\xi}$ corresponds to the adjacency matrix of a weighted Poissonian bipartite random graph with two types
of nodes \cite{Perez2008}: $i$-nodes, associated with the row index of $\bm{\xi}$, and $\mu$-nodes, associated with the column index of $\bm{\xi}$. The total number of nodes is $N + P$ and the average degree of the $\mu$-nodes is $d$, while $c=\alpha d$ is the average degree of the $i$-nodes, with $\alpha=P/N$. Now we introduce the ensemble of $N\times N$ symmetric diluted Wishart random matrices, in which the entries $M_{ij}$ of a given covariance random matrix $\bM$ are obtained from 
\beeq{
  M_{ij}=\frac{1}{d}\sum_{\mu=1}^P \xi_{i\mu}\xi_{j\mu}\,.
  \label{aa2}
}

Defining $\lambda_1,\ldots,\lambda_N$ as the $N$ positive eigenvalues of $\bM$, we are interested here in the statistics of the random variable
\begin{equation}
  \mathcal{I}_N(x) = \sum_{\alpha=1}^N \Theta \left( x - \lambda_{\alpha}  \right)\,,
  \label{jjaqp1}
\end{equation}
which counts the number of eigenvalues smaller than a threshold $x\in\mathbb{R}^{+}$ ($\Theta$ is the Heaviside
function). The main goal here is to study the cumulants of $\mathcal{I}_N(x)$ and the rate function controlling its large deviation probability. For the random matrix model defined by eqs. (\ref{aa1}) and (\ref{aa2}), the main difficulty in pursuing an analytical study of the fluctuations of $\mathcal{I}_N(x)$ lies in the absence of an invariance property of the ensemble and the corresponding unknown analytical form of the joint distribution of eigenvalues \cite{BookMehta}. Therefore, the analytical method of \cite{Vivo} is simply inapplicable. In spite of that, one may still hope to derive analytical results for the statistics of $\mathcal{I}_N(x)$, provided one finds an explicit link between  $\mathcal{I}_N(x)$ and the random matrix  $\bm{M}$. Such link is obtained  by using the representation of the Heaviside function in terms of the discontinuity of the principal value of the complex logarithm
\beeq{
  \mathcal{I}_{N}(x)=\frac{1}{2 \pi i} \lim_{\epsilon\to0^{+}} \sum_{\alpha=1}^N
  \left[ \text{Log}\left( \lambda_\alpha - x_\epsilon   \right) - \text{Log}\left( \lambda_\alpha - \overline{x_\epsilon}   \right) \right]
  \label{ddq1}
}
where $x_{\epsilon}=x-i\epsilon$, and $\overline{(\dots)}$ denotes the complex conjugate.

The above identity is the starting point to study the statistics of $\mathcal{I}_{N}(x)$. The cumulant generating function of  $\mathcal{I}_{N}(x)$ is defined as
\beeq{
\mathcal{F}_x(y)=-\lim_{N\to\infty}\frac{1}{N}\ln \bracket{e^{-y\mathcal{I}_N(x) }}_{\bm{M}}\,,
\label{eq2}
}
in which $\bracket{\dots}_{\bm{M}}$ is the ensemble average with the distribution of the random matrix $\bm{M}$. The  $\ell$-cumulant of $\mathcal{I}_{N}(x)$, defined as
\begin{equation}
\kappa_\ell(x)=\frac{\bracket{\mathcal{I}^\ell_N(x)}^{c}_{\bm{M}}}{N}  
\end{equation}
in terms of the connected correlation  $\bracket{\cdots}^{c}$, is obtained from $\mathcal{F}_x(y)$ according to
\begin{equation}
  \kappa_\ell(x)=(-1)^{\ell+1}\frac{\partial ^\ell \mathcal{F}_x(y)}{\partial y^\ell}\Big|_{y=0} \,.
  \label{cumss}
\end{equation}
Invoking the G\"artner-Ellis theorem \cite{Touchette}, the probability of having $kN$ eigenvalues smaller than $x$ behaves asymptotically for large $N$ as
\begin{equation}
\text{Prob}[\mathcal{I}_N(x)=kN] \asymp e^{-N\Psi_x(k)},
\end{equation}
where the rate function $\Psi_x(k)$ is obtained from $\mathcal{F}_x(y)$ by the Legendre transform:
\beeq{
-\Psi_x(k)=\underset{y\in \mathbb{R}}{\text{inf}}\left[ k y-\mathcal{F}_x(y)\right]\,.
\label{eq:rf}
}
Thus, all cumulants of $\mathcal{I_N}(x)$ and its large deviation probability follow from the analytical expression for $\mathcal{F}_x(y)$, presented in the next section.


\section{Results} 
\label{sectionresults}

The exponential form of $\mathcal{F}_x(y)$ (see eq. (\ref{eq2})), when combined with the identity (\ref{ddq1}), is suitable for the application of the replica approach in order to calculate the ensemble average and the limit $N \rightarrow \infty$ in eq. (\ref{eq2}). We thoroughly explain all steps of such calculation in the appendix \ref{AppA}, while here we only present the final outcome, namely the analytical expression for the cumulant generating function: 
\begin{widetext}
\begin{eqnarray}
  \mathcal{F}_x(y)&=&{\mathcal{A}} \int d\Delta d\Gamma w_{\rho}(\Delta)w_{k}(\Gamma)
  \exp{\left[ -\frac{i y}{2\pi}\text{Log}\left(\frac{\Delta+\frac{1}{\Gamma}}{\overline{\Delta}+\frac{1}{\overline{\Gamma}}}\right) \right] }
  -\alpha\ln \left(\int d\sigma  w_{\sigma}(\sigma )
    \exp{\left[ -\frac{iy}{2\pi}\text{Log}\left(\frac{1+\sigma}{1+\overline{\sigma }}\right) \right]} \right)
  \nonumber \\
  &-&\ln \left( \sum_{\ell=0}^\infty\frac{{\mathcal{A}}^\ell}{\ell!}\int\left( \prod_{s=1}^\ell d\Gamma _s w_{k}(\Gamma_s)\right)
    \exp{\left[\frac{i y}{2\pi} W_{\epsilon}(\Gamma_1,\dots,\Gamma_\ell)  \right]  } \right).
\label{eq:cgf}
\end{eqnarray}
\end{widetext}
The weight $W_{\epsilon}(\Gamma_1,\dots,\Gamma_\ell)$ in the above equation is given by
\begin{equation}
  W_{\epsilon}(\Gamma_1,\dots,\Gamma_\ell)= \sum_{s=1}^\ell\text{Log} \left(\frac{\Gamma_s}{\overline{\Gamma_s}}\right) - \text{Log}\left(\frac{\sum_{s=1}^\ell \Gamma_s-x_\epsilon}{\sum_{s=1}^\ell \overline{\Gamma_s}-\overline{x_\epsilon}}\right),
  \label{wei}
  \end{equation}
while $w_{\rho}(\Delta)$, $w_{k}(\Gamma)$, and $w_{\sigma}(\sigma)$ are the distributions of the complex variables $\Delta$, $\Gamma$ and $\sigma$. These distributions solve the following set of self-consistency equations
\begin{widetext}
\begin{eqnarray}
  w_{\rho}(\Delta)&=&\frac{1}{\mathcal{N}_1}\sum_{\ell=0}^\infty \frac{e^{-{\mathcal{A}}}{\mathcal{A}}^\ell}{\ell!}
  \int \left( \prod_{s=1}^\ell d\Gamma_sw_{k}(\Gamma_s)\right)
 \exp{\left[\frac{i y}{2\pi} W_{\epsilon}(\Gamma_1,\dots,\Gamma_\ell)  \right]  }
  \delta\left(\Delta-\frac{1}{\sum_{s=1}^\ell\Gamma_s-x_\epsilon}\right) , \label{1}\\
  w_{\sigma}(\sigma)&=&\sum_{k=0}^\infty\frac{e^{-d}d^k}{k!}\int\left( \prod_{\ell=1}^kd\Delta_\ell w_{\rho}(\Delta_\ell)d\xi_\ell P_\xi(\xi_\ell)\right)
  \delta\left( \sigma - \frac{1}{d}\sum_{\ell=1}^k\xi^2_\ell \Delta_{\ell}\right) , \label{2}\\
w_{k}(\Gamma)&=&\int d\xi P_\xi(\xi) \int d\sigma w_{\sigma}(\sigma)
\delta\left[ \Gamma-\frac{\xi^2}{d \left( 1+\sigma \right) } \right] .  \label{3}
\end{eqnarray}
\end{widetext}
The constant factors $\mathcal{N}_1$ and $\mathcal{A}$, appearing in eqs. (\ref{eq:cgf}) and (\ref{1}), are
defined as follows
\begin{eqnarray}
  \mathcal{N}_1&=&\sum_{\ell=0}^\infty \frac{e^{-{\mathcal{A}}}{\mathcal{A}}^\ell}{\ell!}
  \int \left( \prod_{s=1}^\ell d\Gamma_sw_{k}(\Gamma_s) \right) \nonumber \\
 &\times& \exp{\left[\frac{i y}{2\pi} W_{\epsilon}(\Gamma_1,\dots,\Gamma_\ell)  \right]  }  \,,  \\
 {\mathcal{A}}&=&\frac{\alpha d }{\int d\sigma w_{\sigma}(\sigma)
   \exp{ \left[ -\frac{i y}{2 \pi}\text{Log}\left(\frac{1+\sigma}{1+\overline{\sigma}}\right)  \right] }  }\,. \label{eq:A}
\end{eqnarray}
Equation (\ref{eq:cgf}) is the main analytical result of our work, since it provides the exact  cumulant generating function of $\mathcal{I}_N(x)$ for the sparse Wishart ensemble in the limit $N \rightarrow \infty$, from which the rate function and the behavior of all cumulants readily follow. Notice that $\mathcal{F}_x(y)$ depends on the distributions $w_{\rho}(\Delta)$, $w_{k}(\Gamma)$ and $w_{\sigma}(\sigma)$ through the solution of eqs. (\ref{1}-\ref{3}), for which there is no closed, analytical form in the general case. Hence we must solve eqs. (\ref{1}-\ref{3}) numerically, which can be done very efficiently using the weighted population dynamics algorithm. All steps of this numerical method are carefully discussed in appendix \ref{Weighted}. We point out that the limit $\epsilon \rightarrow 0^+$ is implicitly assumed in eqs. (\ref{eq:cgf}-\ref{eq:A}).

Equation \eqref{eq:cgf} can be interpreted as a combination of effective moments or cumulant generating functions of the following random  variables
\begin{align}
  \mathcal{I}_1(\Delta,\Gamma) &\equiv \frac{i}{2\pi}
  \text{Log} \left( \frac{\frac{1}{\overline{\Gamma}}+\overline{\Delta}}{\frac{1}{\Gamma}+\Delta} \right)\,, \label{11} \\
\mathcal{I}_2(\sigma) &\equiv \frac{i}{2\pi} \text{Log} \left( \frac{1+\overline{\sigma}}{1+\sigma} \right)\,, \label{12} \\
\mathcal{I}_3(\{\Gamma_s\}_{s=1}^\ell) &\equiv \frac{i}{2\pi}
W_{\epsilon}(\Gamma_1,\dots,\Gamma_\ell) . \label{13}
\end{align}
From eq. (\ref{eq:cgf}), we note that the dependence of  $\mathcal{F}_x(y)$ on $y$ is explicit, as the generating parameter, and implicit, since the measures $w_\rho$, $w_k$, and $w_{\sigma}$ do also depend on $y$ through the saddle-point eqs. (\ref{1}-\ref{3}). However, it can be shown that, for the first two cumulants, the implicit derivative is not needed and one readily obtains
\begin{eqnarray}
  \kappa_1(x)&=&-\alpha d\bracket{\mathcal{I}_1(\Delta,\Gamma)} + \bracket{\mathcal{I}_3(\{\Gamma_s\}_{s=1}^\ell)} \nonumber \\
  &+& \alpha \bracket{\mathcal{I}_2(\sigma)},\label{sddwewe} \\
  \kappa_2(x)&=&\alpha d\bracket{\mathcal{I}^2_1(\Delta,\Gamma)} \nonumber  \\
  &-& [\bracket{\mathcal{I}^2_3 \left(\{\Gamma_s\}_{s=1}^\ell \right)}
    -\bracket{\mathcal{I}_3 \left( \{\Gamma_s\}_{s=1}^\ell \right)}^2]  \nonumber \\
    &-& \alpha\left[ \bracket{\mathcal{I}^2_2(\sigma)} - \bracket{\mathcal{I}_2(\sigma)}^2\right]\,,
    \label{dd1}
\end{eqnarray}
in which the brackets $\langle \dots   \rangle$ denote the average over the complex random variables $\Delta$, $\Gamma$ and $\sigma$ with their corresponding distributions evaluated at $y=0$. This average is calculated through the population dynamics method, as explained in appendix \ref{Weighted}. Although we do not derive here the analytical equations for higher order cumulants, in principle one can study their large-$N$ behavior by computing numerically the derivatives of eq. (\ref{eq:cgf}) with respect to $y$. 

Let us now present explicit numerical results obtained from our theoretical approach. Although our equations are valid for arbitrary $\alpha=P/N$, here we limit ourselves to the regime $\alpha > 1$, in which the covariance random matrix is nonsingular and, consequently, more relevant for practical applications. For the sake of simplicity, we present results only for the case $P_\xi(\xi) = \delta(\xi-1)$.

\begin{figure}[h]
  \subfigure[$\, \alpha=2$ and $d=1$.]{
    \includegraphics[scale=0.46]{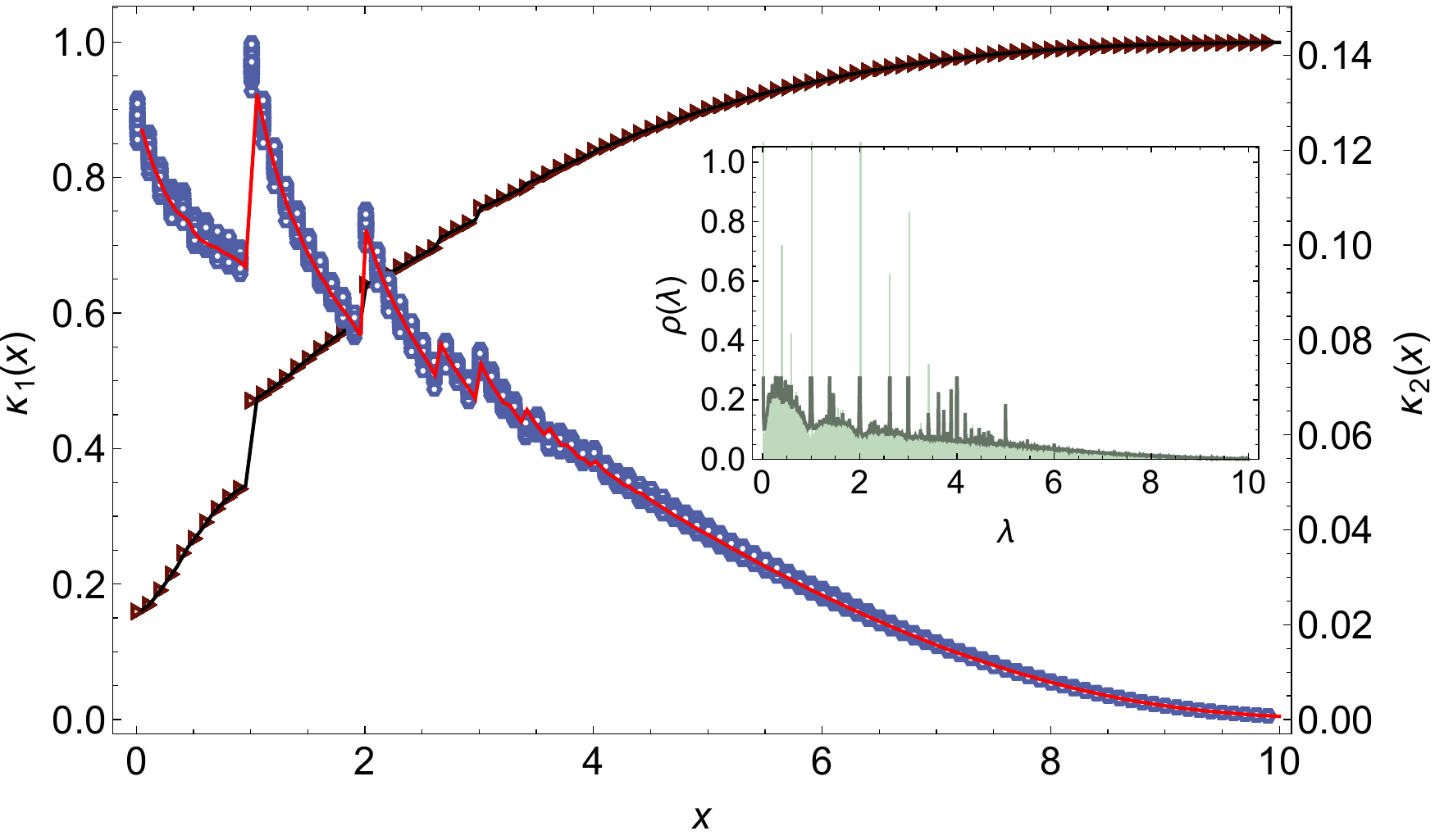}
\label{fig1a}
  }
   \subfigure[$\, \alpha = 2$ and $d=2$.]{
     \includegraphics[scale=0.46]{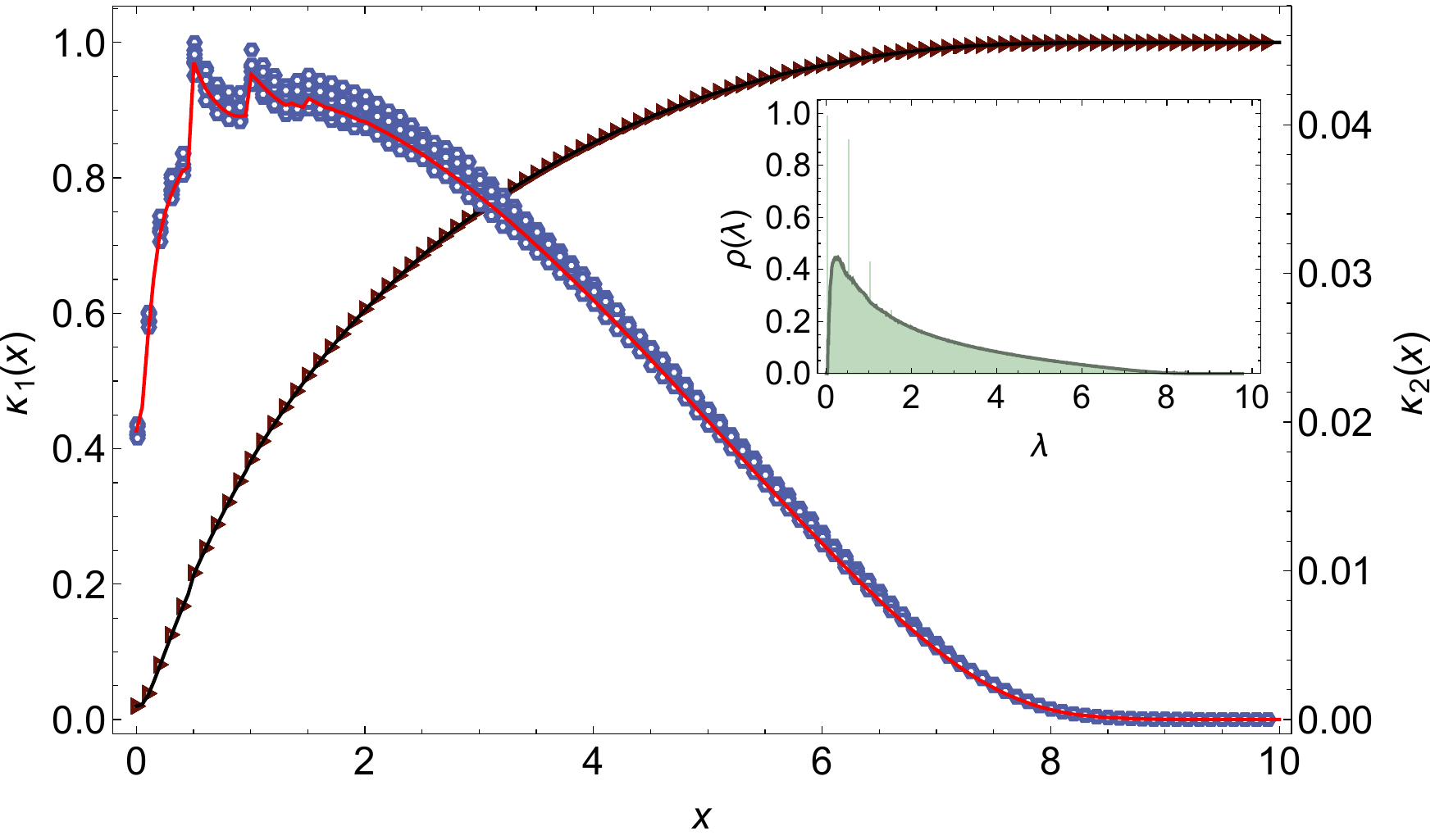}
\label{fig1b}
   }
\caption{The first and second cumulants of $\mathcal{I}_N(x)$ as functions of $x$ for different values of $\alpha$ and $d$. The solid lines are obtained from the solution of eqs. (\ref{1}-\ref{3}) using the weighted population dynamics method with $\mathcal{L}= 10^7$ and $\epsilon = 10^{-8}$ (see appendix \ref{Weighted}). The black curve is the mean value of $\mathcal{I}_N(x)$, whereas the red curve gives the variance of $\mathcal{I}_N(x)$.The markers correspond to results coming from numerical diagonalization of an ensemble with $10^{4}$ sparse Wishart matrices of dimension $N=1000$. The insets display the average spectral density $\rho(\lambda)$ for each pair of model parameters.}
\label{fig1}
\end{figure}

In figure \ref{fig1} we illustrate the behavior of the first two cumulants of $\mathcal{I}_N(x)$ and compare the outcome of our theory with direct diagonalization of finite random matrices. Since the diluted Wishart ensemble is characterized by two parameters $(\alpha,d)$, we have chosen in figure \ref{fig1} the values $(\alpha,d)=(2,1)$ and $(\alpha,d)=(2,2)$, as the spectral density displays rather distinct features for these two pairs of values: while in the first case the continuous part of the spectrum is bathed in multiple Dirac delta peaks, in the second case the contribution of the discrete eigenvalues is less important. The existence of delta peaks is a typical feature in the spectral properties of sparse random matrices \cite{Perez2008,Kuhn2008, MetzA2010}. The peaks are commonly located at the eigenvalues of isolated, disconnected trees, that appear for small average degrees of the associated random graph model \cite{Bauer2001}. As clearly shown in figure \ref{fig1a}, the presence of delta peaks in the spectrum manifests itself as discontinuities in the behavior of the first two cumulants, similarly to the analogous results derived for the adjacency matrix of random graphs \cite{Metz2015,Metz2016}.

\begin{figure}[h]
\includegraphics[scale=0.65]{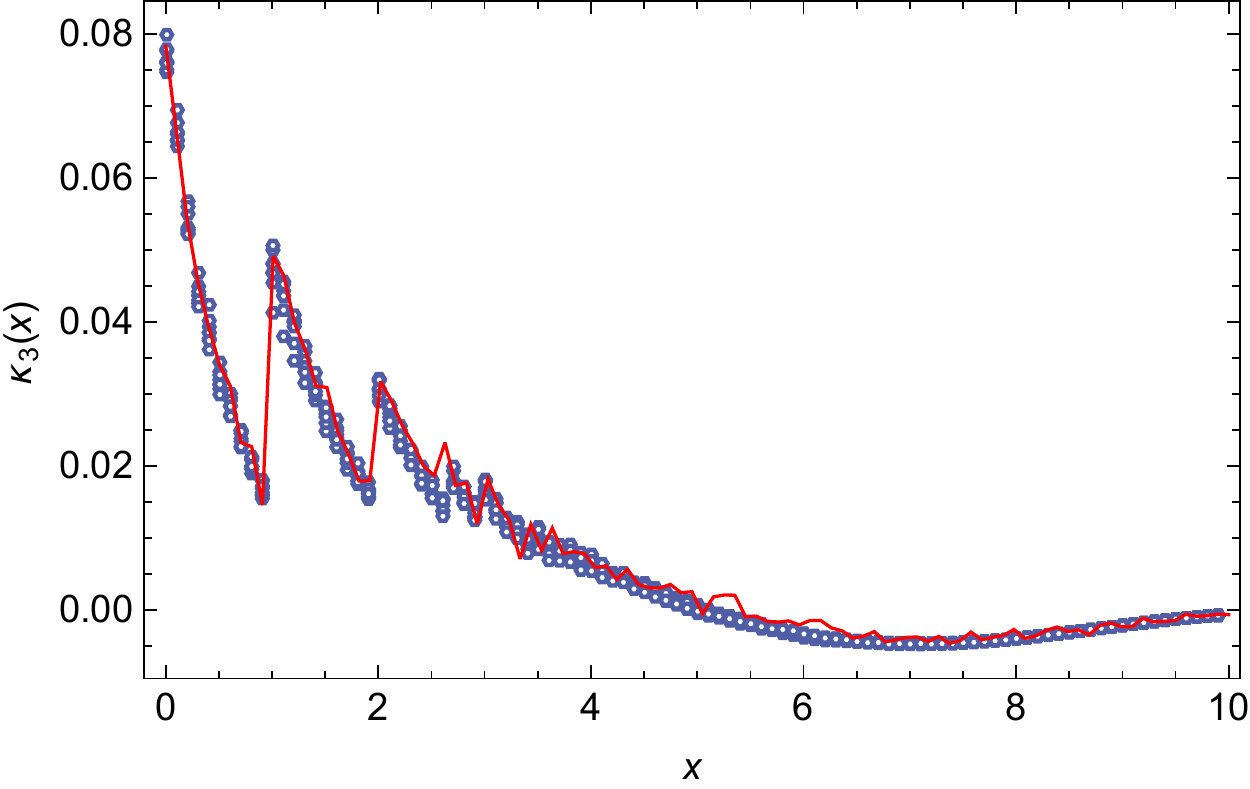}
\caption{The third  cumulant of $\mathcal{I}_N(x)$ as a function of $x$ for $(\alpha,d)=(2,1)$. The solid line has been obtained by evaluating the cumulant generating function $\mathcal{F}_x(y)$ from eq. (\ref{eq:cgf}) at seven points of $y$, followed by a finite differences calculation of the third-order derivative of $\mathcal{F}_x(y)$. These results have been computed through the weighted population dynamics algorithm with $\mathcal{L}= 10^6$ and $\epsilon = 10^{-8}$. Moreover, each estimate of $\mathcal{F}_x(y)$ has been averaged a hundred times in order to reduce its fluctuations and improve the accuracy of the finite differences calculation. The markers correspond to results coming from numerical diagonalization of an ensemble with $10^{6}$ sparse Wishart matrices of dimension $N=300$. }
\label{fig3}
\end{figure}

In order to further inspect the validity of our theory, we have also compared the third cumulant of $\mathcal{I}_N(x)$, derived from eq. (\ref{eq:cgf}), with direct diagonalization. In this case, we need to take into account the explicit and implicit dependences of the cumulant generating function with respect to $y$. Instead of trying to derive a set of self-consistency equations for the derivatives of the distributions $w_{\rho}$, $w_{\sigma}$ and $w_{k}$, we have opted to  evaluate the third derivative of $\mathcal{F}_x(y)$ by finite differences, using the algorithm presented in \cite{Fornberg1988}. In figure  \ref{fig3} we show the behavior of the third cumulant $\kappa_3(x)$ for $(\alpha,d)=(2,1)$. Apart from small fluctuations, the agreement between our theoretical approach and numerical diagonalization is very good.

\begin{figure}[h]
  \subfigure[$\, \alpha=2$ and $d=1$.]{
    \includegraphics[scale=0.65]{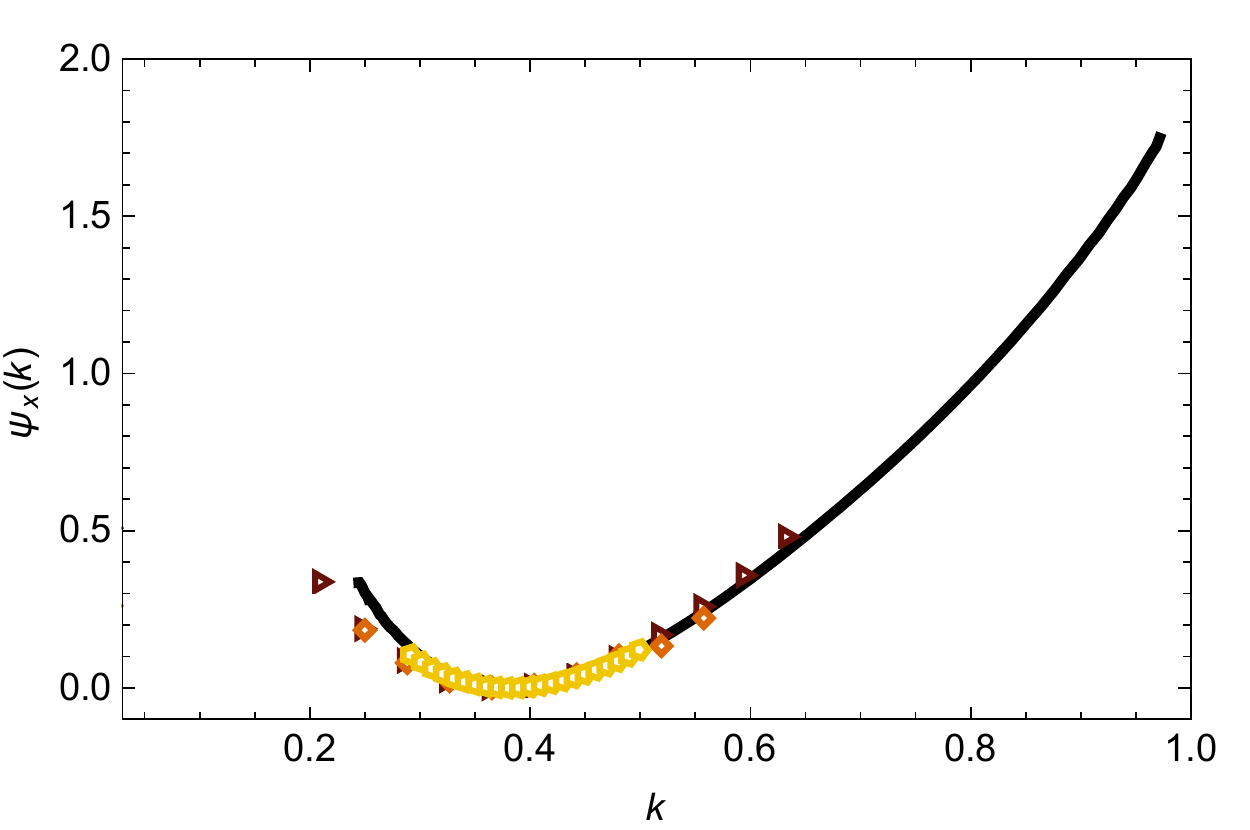}
\label{fig1aaa}
  }
   \subfigure[$\, \alpha=2$ and $d=2$.]{
     \includegraphics[scale=0.65]{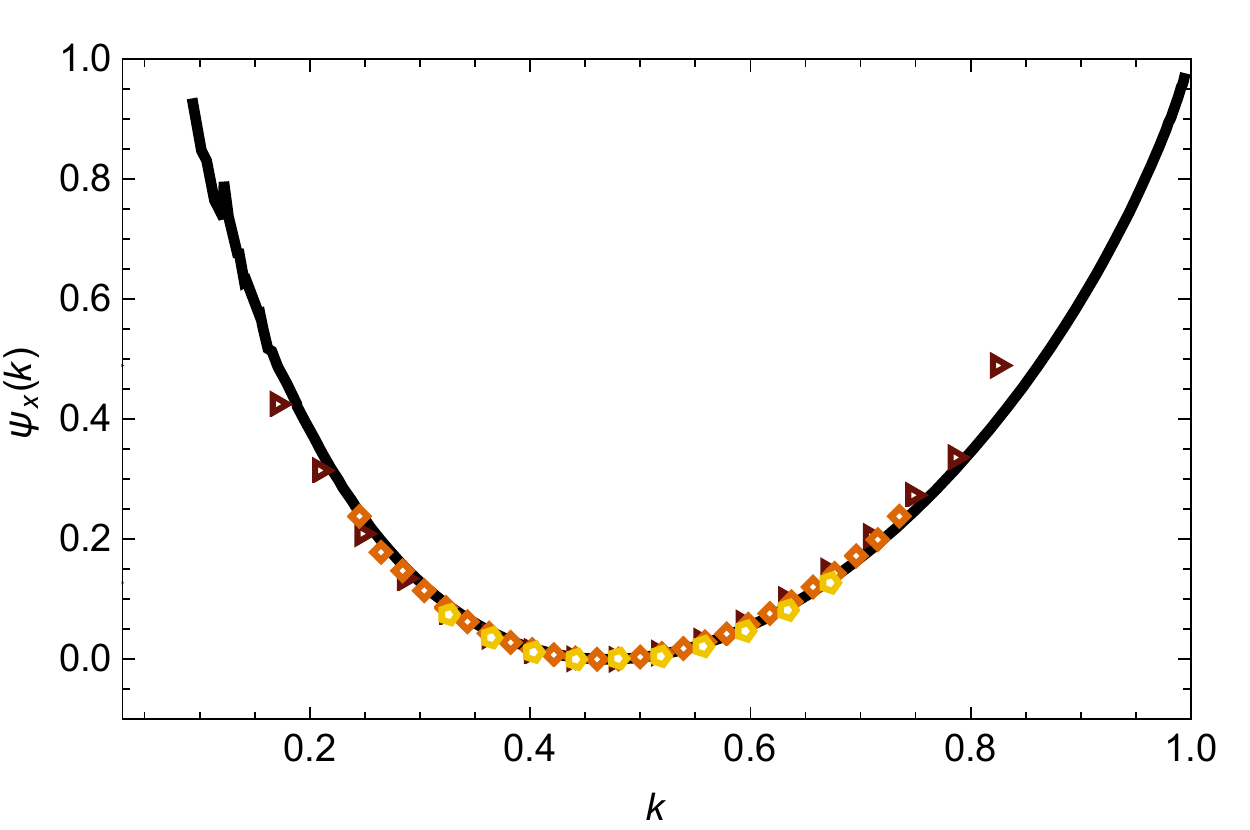}
\label{fig1aab}
   }
\caption{The rate function of the fraction of eigenvalues $k$ inside the interval $(-\infty,x]$ for $x=1.01$ and two pairs of parameters characterizing the sparse Wishart ensemble. The solid black line is derived from the solution of eqs. (\ref{1}-\ref{3}) using the weighted population dynamics method with $\mathcal{L}= 10^6$ and $\epsilon = 10^{-8}$ (see appendix \ref{Weighted}). The markers are results obtained from numerical diagonalization of an ensemble with $10^{6}$ sparse Wishart matrices of dimensions $N=25$ (brown), $N=50$ (orange) and $N=100$ (yellow).}
\label{fig1aa}
\end{figure}

Let us now turn our attention to the behavior of the rate function $\Psi_{x}(k)$. Figure \ref{fig1aa} illustrates the rate function for $x=1.01$ and the same combinations of model parameters as discussed in figure \ref{fig1}: $(\alpha,d)=(2,1)$ and $(\alpha,d)=(2,2)$. The outcome of our theory is compared with direct diagonalization of finite random matrices, showing an excellent agreement in the range of values of $k$ that can be probed through numerical diagonalization. Indeed, the probability of observing $\mathcal{I}_N(x) = kN$ decays exponentially with $N$, which promptly hinders any attempt to explore a sizeable interval of $k$ for large $N$, as one needs to diagonalize an unfeasible number of finite random matrices. This remarkable limitation of numerical diagonalization procedures further illustrate the importance of our theory, since we can virtually determine the rate function $\Psi_{x}(k)$ for any value of $0 \leq k \leq 1$.

As can be noted in figure \ref{fig1aa}, the function $\Psi_{x}(k)$ is asymmetric around its minimum, whose value of $k$ coincides with the first cumulant of $\mathcal{I}_N(x)$, calculated from eq. (\ref{sddwewe}). Such asymmetric feature is strikingly distinct from the symmetry of the analogous rate function in the classical Wishart ensemble \cite{Vivo}. This seems to be a general property of sparse random matrix ensembles \cite{Metz2016}, essentially due to the presence of Dirac delta peaks in their spectral density \cite{Bauer2001}. Concerning figure \ref{fig1aaa}, we conclude that random matrix samples that increase $\mathcal{I}_N(x)$ are more likely, since the rate function grows slowly for $k$ larger than its typical value. This is due to the presence of a Dirac delta peak with a large weight precisely at $x=1$, as clearly shown by the discontinuous behavior of $\kappa_1(x)$ at $x=1$ (see figure \ref{fig1a}). One expects that the weights of the discrete contributions to the spectral density decrease exponentially with the average degree $d$ \cite{Bauer2001}, which results in a more symmetric rate function for larger values of $d$. This picture is consistent with the results in figures \ref{fig1} and \ref{fig1aa}.

\section{Conclusion}

We have discussed here a theoretical approach to study the eigenvalue fluctuations of sparse Wishart random matrices, in which the members of the random matrix ensemble are covariance matrices with a large amount of zero entries. Such class of random matrices is of fundamental importance in various techniques to process large datasets in multivariate statistics. The main outcome of our work is an analytical expression (see eq. (\ref{eq:cgf})) for the $N \rightarrow \infty$ behavior of the cumulant generating function of the number of eigenvalues $\mathcal{I}_N(x)$  within the interval $(-\infty,x]$. This analytical formula is the main source of quantitative information about the fluctuations of $\mathcal{I}_N(x)$. In fact, we have presented explicit results for the mean and the variance of  $\mathcal{I}_N(x)$, its rate function characterizing the probability of rare events, and the third-order cumulant of  $\mathcal{I}_N(x)$.

  Similarly to previous results for the adjacency matrix of random graphs \cite{Metz2016}, the rate function controlling the large deviation probability of $\mathcal{I}_N(x)$ is asymmetric in the case of the sparse Wishart ensemble. This feature is particularly evident in the regime of high sparseness, where delta peaks due to isolated clusters in the associated random graph are abundant in the spectral density. It is reasonable to expect that the existence of Dirac delta peaks with large weights, at which a large number of eigenvalues accumulates, strongly influences the typical and rare fluctuations of  $\mathcal{I}_N(x)$, being responsible for the asymmetry of the rate function. It would be interesting to further explore this issue by explicitly disentangling the contributions to the rate function coming from the discrete and continuous parts of the spectral density \cite{Kuhn2016}.

  Finally, the present work further reinforces the exactness of the powerful method designed in \cite{Metz2016}. In order to test the validity of our analytical results, we have carefully compared them with numerical diagonalization of large random matrices and the agreement between these two independent approaches is very good. This further corroborates our theoretical method, in spite of its lack of a full mathematical rigor (see appendix \ref{AppA}).


\appendix

\section{Derivation of the cumulant generating function} \label{AppA}
The number of eigenvalues $\mathcal{I}_{N}(x)$ within the unbounded interval $(-\infty,x]$, defined in eq. (\ref{jjaqp1}), can be rewritten using an identity for the Heaviside function \cite{Metz2015,Metz2016}
\begin{eqnarray}
  \mathcal{I}_{N}(x)=\lim_{\epsilon\to0^{+}}\frac{1}{2\pi i} \sum_{\alpha=1}^N \Big[ \text{Log} (\lambda_{\alpha}+i\epsilon-x) \nonumber \\
    - \text{Log} (\lambda_{\alpha}-i\epsilon-x)\Big]\,,
\label{eq3}
\end{eqnarray}
where $\text{Log}(\dots)$ corresponds to the principal branch of the complex logarithm. One of the central formulas in applying spin-glass techniques to random matrix theory is the following result for the multi-dimensional Fresnel integral
\begin{align}
  Z(x_\epsilon)&=\int \left(  \prod_{i=1}^N  d y_i \right) \exp\left[-\frac{i}{2}\bm{y}^T.(x_\epsilon\mathbb{I}-\bm{M})\bm{y}\right]\nonumber \\
  &=(2\pi)^{N/2}\exp\left[-\frac{1}{2}\sum_{i=1}^N\text{Log} (\lambda_i - x_\epsilon)+i\frac{N\pi}{4} \right],
\label{eq:fir}
\end{align}
where $x_{\epsilon}=x-i\epsilon$, $\bm{y}^T=(y_1,\ldots, y_N)$, and $\mathbb{I}$ is the $N \times N$ identity matrix. In the complex plane, neither the exponential nor the logarithm are injective functions. While this is not a problem when studying the spectral density,  some initial caution is warranted when considering the statistics of $\mathcal{I}_{N}(x)$. Being na\"ive with these multivalued functions in the complex plane, eq. (\ref{eq3}) can be written as follows
\beeq{
\mathcal{I}_{N}(x)=\lim_{\epsilon\to0^{+}}\frac{1}{\pi i}\Bigg[\text{Log}\overline{Z(x_\epsilon)}-\text{Log}Z(x_\epsilon)+\frac{N\pi i}{2}\Bigg]
\label{eq2ab}\,,
}
with $\overline{(\dots)}$ denoting complex conjugation.

\begin{figure}[H]
\includegraphics[scale=0.62]{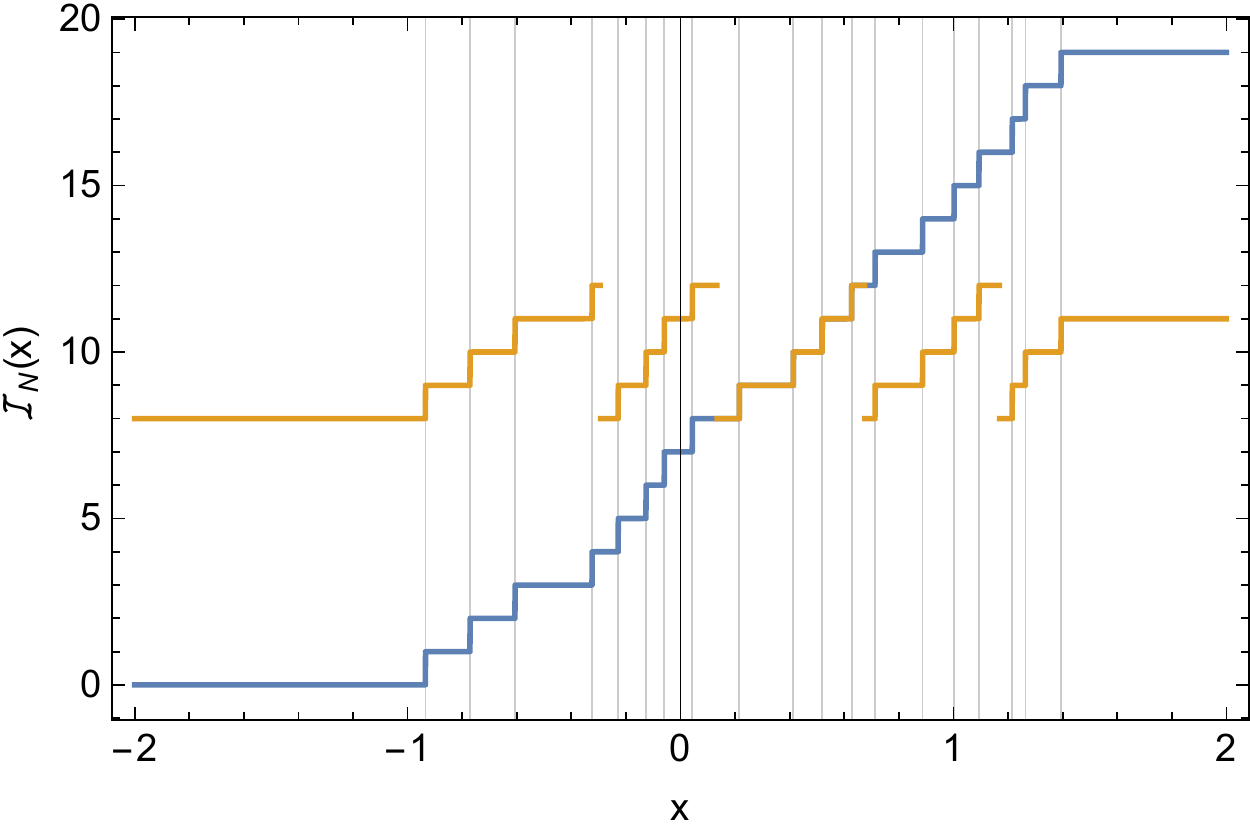}
\caption{The shifted index number $\mathcal{I}_{N}(x)$ for a single realization of an $N \times N$ random matrix. The function
  $\mathcal{I}_{N}(x)$ counts the number of eigenvalues smaller than $x$. Here we show a comparison between two
  representations of $\mathcal{I}_{N}(x)$: eq. \eqref{eq2ab} (solid orange line) and eq. \eqref{eq3} (solid blue line). The vertical
  grey lines denote the positions of each one of the eigenvalues $\{\lambda_{\alpha}\}_{\alpha=1}^{20}$.}
\label{fig1SI}
\end{figure}

Clearly, expression \eqref{eq2ab} does not provide the correct result for $\mathcal{I}_{N}(x)$, given precisely by eq. \eqref{eq3}. This situation is illustrated in Figure  \ref{fig1SI}, where we compare the outcomes of eqs. \eqref{eq2ab} and \eqref{eq3} for  a single random matrix of size $N=20$. Thus, eq. \eqref{eq2ab} is unfit to count the number of eigenvalues for a single realization of $\bm{M}$. The reason is that we have na\"ively folded in the sum appearing in eq. \eqref{eq3} as products inside the logarithms in eq. \eqref{eq2ab}, without any regard of the multivalued character of the complex logarithm. In other words, we have assumed that the complex logarithm fulfills the same standard properties as  the logarithm of a real variable.

Although eq. (\ref{eq2ab}) is unsuitable to count the number of eigenvalues, it is the appropriate starting point to compute the ensemble average using the replica approach. It is thus a relevant question whether the above procedure leads to correct results for the statistics of the index. As we show in the paper, our method does yield correct results when comparing the final analytical expressions for the  rate function and for the cumulants of $\mathcal{I}_{N}(x)$ with numerical diagonalization, which shows that  eq. \eqref{eq2ab} correctly encodes the statistical properties of the index. We have some hint as to why this naive procedure works: we call it the {\it folding/unfolding mechanism}. After going from eq. \eqref{eq3} to eq. \eqref{eq2ab}, one derives an effective theory using the replica method, which essentially consists of decoupling sites by coupling replicas \cite{BookParisi}. This theory  can then be unfolded by factorizing over the sites through the introduction of a suitable order-parameter, restoring the correct scaling of the moments of $\mathcal{I}_{N}(x)$ with respect to $N$. The non-injective nature of the logarithm in the complex plane seems irrelevant for the statistical properties of $\mathcal{I}_{N}(x)$, which implies that we can be careless about the actual prescription used to represent the multidimensional Gaussian integral.

Let us now proceed to the analytical computation of the cumulant generating function. By substituting eq. (\ref{eq2ab}) in eq. (\ref{eq2}), the cumulant generating function can be expressed as follows
\beeq{
  \mathcal{F}_x(y)&=\frac{y}{2}-\lim_{N\to\infty} \lim_{\epsilon\to 0^+}  \frac{1}{N}\ln \bracket{ [Z(x_\epsilon)]^{\frac{iy}{\pi }}[\overline{Z(x_\epsilon)}]^{-\frac{iy}{\pi }} }_{\bm{M}} ,
}
with $\langle \dots  \rangle_{\bM}$ denoting the ensemble average over the distribution of $\bM$ (see eqs. (\ref{aa1}) and (\ref{aa2})).
In order to calculate  $\mathcal{F}_x(y)$  for large $N$, we use the replica method in the form of the identity
\beeq{
  \mathcal{F}_x(y)&=\frac{y}{2}- \lim_{N\to\infty} \lim_{\epsilon\to 0^+} \lim_{n_{\pm} \to \pm \frac{iy}{\pi }  }
  \frac{1}{N}
  \ln  \mathcal{Q}_N(n_{\pm},x_\epsilon),
  \label{hhaap}
}
where we have introduced the object
\beeq{
 \mathcal{Q}_N(n_{\pm},x_\epsilon)=   \bracket{[Z(x_\epsilon)]^{n_{+}}[\overline{Z(x_\epsilon)}]^{n_-}}_{\bm{M}}.
}
Initially, the exponents  $n_{\pm}$  are assumed to be positive integers. After the ensemble average in the above equation has been calculated for $N \rightarrow \infty$, we perform the replica limit $n_{\pm}\to\pm\frac{i y}{\pi}$ and make an analytical continuation of $n_{\pm}$ to the imaginary axis.

Using the Fresnel integral representation of eq. \eqref{eq:fir}, we write $\mathcal{Q}_N(n_{\pm},x_\epsilon)$ as follows
\begin{align}
  \mathcal{Q}_N(n_{\pm},x_\epsilon)&= \int \left(\prod_{a=1}^{n_{+}}d\bm{z}_{a}\prod_{a=1}^{n_{-}}d\bm{y}_{a}\right) \nonumber \\
  &\times
  \exp\left[-\frac{i x_\epsilon}{2}\sum_{a=1}^{n_{+}}\sum_{i=1}^Nz^2_{ia}+\frac{i \overline{x_\epsilon}}{2}\sum_{a=1}^{n_{-}}\sum_{i=1}^Ny_{ia}^2\right] \nonumber \\
&\times
  \bracket{\exp\left[\frac{i}{2}\sum_{i,j=1}^NM_{ij}
      \left(\underline{z}_{i}\cdot \underline{z}_{j}-\underline{y}_{i}\cdot \underline{y}_{j}\right)\right]}_{M},
  \label{jja}
\end{align}
in which $\bm{z}_a\equiv (z_{1a},\ldots z_{Na})$, $\bm{y}_a\equiv (y_{1a},\ldots y_{Na})$, $\underline{z}_i=(z_{i1},\ldots, z_{i n_{+}})$ and $\underline{y}_i=(y_{i1},\ldots, y_{i n_{-}})$, with $i,j=1,\ldots,N$. Equation (\ref{jja}) holds in general for a real symmetric matrix $\bM$. After computing the ensemble average in eq. (\ref{jja}) with the distribution of $\bM$, defined by eqs. (\ref{aa1}) and (\ref{aa2}), and introducing the order-parameter function
\begin{equation}
  \rho(\underline{z},\underline{y}) = \frac{1}{N} \sum_{i=1}^{N} \delta\left(\underline{z} -\underline{z}_i    \right)
  \delta\left(\underline{y} -\underline{y}_i    \right)\,,
  \end{equation}
the quantity $\mathcal{Q}_N(n_{\pm},x_\epsilon)$ can be expressed as a path integral over two functions, $\rho(\underline{z},\underline{y})$ and $k(\underline{z},\underline{y})$, defined on the replica space $(\underline{z},\underline{y})\in\mathbb{R}^{n_+}\times\mathbb{R}^{n_-}$
\begin{equation}
  \mathcal{Q}_N(n_{\pm},x_\epsilon) = \int D[\rho,k]e^{-N\mathcal{S}[\rho,k]}\,, \label{gha} 
\end{equation}
where the action reads
  \begin{align}
    \mathcal{S}[\rho,k]&\equiv -i \int d\underline{z} d\underline{y} k(\underline{z},\underline{y})\rho(\underline{z},\underline{y}) \nonumber \\
    &-\ln  \left[ \int d\underline{z}d\underline{y}\exp\left(-\frac{ix_\epsilon}{2}\underline{z}^2+\frac{i\overline{x_\epsilon}}{2}\underline{y}^2
    -ik(\underline{z},\underline{y})\right) \right] \nonumber \\
  &-\alpha \ln\int D[\underline{m},\underline{L}]e^{\frac{i}{2} \left(\underline{m}^2-\underline{L}^2\right)}
    \mathcal{M}(\underline{m},\underline{L}|\rho) \,, \label{action}
\end{align}
  with
  \begin{align}
    \mathcal{M}&(\underline{m},\underline{L}|\rho) \equiv \sum_{k=0}^\infty\frac{e^{-d}d^k}{k!} \nonumber \\
    &\times
    \int\left(\prod_{\ell=1}^k d \underline{z}_\ell
    d\underline{y}_\ell \rho(\underline{z}_\ell,\underline{y}_\ell)d\xi_\ell P_\xi(\xi_\ell)\right) \nonumber \\
&\times
    \delta\left( \underline{m}-\frac{1}{\sqrt{d}}\sum_{\ell=1}^k \xi_\ell \underline{z}_\ell\right)\delta \left(\underline{L}-\frac{1}{\sqrt{d}}\sum_{\ell=1}^k \xi_\ell \underline{y}_\ell\right). \label{jjaqp}
\end{align}
The quantity $k(\underline{z},\underline{y})$ is the function conjugate to the order-parameter $\rho(\underline{z},\underline{y})$. The next step consists in evaluating the path integral by applying the saddle-point method, which captures the leading contribution to the integral in the limit $N \rightarrow \infty$. Plugging eq. (\ref{gha}) in eq. (\ref{hhaap}) and taking the limit $N \rightarrow \infty$, we obtain
\beeq{
\mathcal{F}_x(y)&= \frac{y}{2} + \lim_{\epsilon\to 0^+} \lim_{n_{\pm} \to \pm \frac{iy}{\pi }}  \mathcal{S}_0[\rho,k]\,,
\label{eq:abc}
}
where $ \mathcal{S}_0[\rho,k]$ is the action $S[\rho,k]$ evaluated at its saddle-point, at which the functions $\rho(\underline{z},\underline{y})$ and $k(\underline{z},\underline{y})$ obey the stationary equations:
\beeq{
\frac{\delta S[\rho,k]}{\delta \rho(\underline{z},\underline{y})}= \frac{\delta S[\rho,k]}{\delta k(\underline{z},\underline{y})}=0.
}
\begin{widetext}
After some straightforward algebra, the saddle-point equations take the following form:
\begin{eqnarray}
k (\underline{z},\underline{y})&=& i \alpha d \int d\xi P_\xi(\xi)\frac{\int D[\underline{m},\underline{L}]e^{\frac{i}{2}\left(\underline{m}^2- \underline{L}^2\right)} \mathcal{M}\left(\underline{m}-\xi \underline{z}/\sqrt{d},L- \xi \underline{y}/\sqrt{d}\big|\rho\right)  }{\int D[\underline{m},\underline{L}]e^{\frac{i}{2} \left(\underline{m}^2- \underline{L}^2\right)}\mathcal{M}(\underline{m},\underline{L}|\rho)},  \label{hhaq}  \\
\rho(\underline{z},\underline{y})&=& \frac{ \exp{\left[-\frac{ix_\epsilon}{2}\underline{z}^2+\frac{i\overline{ x_\epsilon}}{2}\underline{y}^2-ik(\underline{z},\underline{y}) \right]}}{\int d\underline{z}d\underline{y} \exp{\left[-\frac{i x_\epsilon}{2}\underline{x}^2+\frac{i\overline{ x_\epsilon}}{2}\underline{y}^2-ik(\underline{z},\underline{y})\right]}}.
\label{eq:spf}
\end{eqnarray}
\end{widetext}
The replica limit $n_{\pm}\to \pm\frac{iy}{\pi}$ in the saddle-point equations and in the expression \eqref{eq:abc} will be dealt with by assuming that  $\rho(\underline{z},\underline{y})$ and $k(\underline{z},\underline{y})$ are replica symmetric (RS) in the subspaces $\mathbb{R}^{n_{\pm}}$.
\subsection{Replica symmetric {\it ansatz}}
Even though it is possible to express eq. (\ref{action}) solely in terms of the order-parameter function $\rho(\underline{z},\underline{y})$, in the  following derivation we introduce, for the sake of clarity, a replica symmetric {\it ansatz} for each function $k(\underline{z},\underline{y})$, $\rho(\underline{z},\underline{y})$ and $\mathcal{M}(\underline{m},\underline{L}|\rho)$. As shown in previous works dealing with the spectral properties of random graphs \cite{Perez2008,Kuhn2008,Metz2016}, the parametrization of the order-paramater in terms of a superposition of Gaussian functions leads to exact results. Thus, with a modest amount of foresight, we propose the {\it ansatze}
\begin{align}
  \mathcal{M}(\underline{m},\underline{L}|\rho)&= \int d\sigma  w_{\sigma}(\sigma)\left[ \prod_{a=1}^{n_{+}}
    \frac{ \exp{\left( i\frac{m_a^2}{2\sigma} \right) } }{\sqrt{2\pi i \sigma   } }\right] \nonumber \\
  &\times \left[\prod_{a=1}^{n_{-}}\frac{ \exp{\left( -i\frac{L_a^2}{2 \overline{\sigma} } \right)}}{\sqrt{- 2\pi i  \overline{\sigma}  }}\right] , \label{jka1}
\end{align}
\begin{align}
  \rho(\underline{z},\underline{y})&=\int d \Delta  w_{\rho}(\Delta)
  \left[\prod_{a=1}^{n_{+}}\frac{ \exp{\left( i\frac{z_a^2}{2\Delta} \right) } }{\sqrt{2\pi i\Delta  } }\right] \nonumber \\
  &\times \left[\prod_{a=1}^{n_{-}}\frac{ \exp{\left( -i\frac{y_a^2}{2\overline{\Delta}} \right) } }{\sqrt{-2\pi i\overline{\Delta}  } }\right] , \label{jka2}
\end{align}
\begin{align}
  k(\underline{z},\underline{y})&= i \mathcal{A}\int d \Gamma w_{k}(\Gamma)
  \left[\prod_{a=1}^{n_{+}}\sqrt{\frac{\Gamma}{2 \pi i} } \exp{\left( i\Gamma \frac{z_a^2}{2} \right) } \right] \nonumber \\
  &\times \left[\prod_{a=1}^{n_{-}}\sqrt{\frac{\overline{\Gamma}}{-2 \pi i} } \exp{\left( -i \overline{\Gamma} \frac{y_a^2}{2} \right) } \right] , \label{jka3}
\end{align}
It is important to keep track of the normalization factors, that is why we have written them explicitly in the above equations. We also assume that the distributions $w_{\rho}(\Delta)$, $w_{k}(\Delta)$ and  $ w_{\sigma}(\sigma)$ are normalized. Equations (\ref{jka1}-\ref{jka3})  remain invariant under the interchange of replica indexes within each subspace. Inserting the above Gaussian assumptions in eqs. (\ref{jjaqp}), (\ref{hhaq}) and (\ref{eq:spf}), and performing the replica limit $n_{\pm} \rightarrow \pm \frac{i y}{\pi}$, one arrives at the set of self-consistent eqs. (\ref{1}-\ref{3}). The expression for the cumulant generating function, explicitly shown in eq. (\ref{eq:cgf}), is derived by substituting eqs. (\ref{jka1}-\ref{jka3}) in eq. (\ref{eq:abc}) and then taking the limit $n_{\pm} \rightarrow \pm \frac{i y}{\pi}$.

The limit $\epsilon\to 0^{+}$ is implicit in eq. (\ref{eq:cgf}) as well as in the self-consistent eqs. (\ref{1}-\ref{3}). From a mathematical viewpoint, this limit corresponds to recovering a Dirac delta distribution from a Cauchy distribution \cite{Perez2008,Kuhn2008}. However, as we will see below, choosing a small value of $\epsilon$ is enough to obtain excellent numerical results, and the actual limit does not need to be taken.

 \section{Weighted Population dynamics}  \label{Weighted}

 In order to solve  numerically eqs. (\ref{1}-\ref{3}), we use the weighted population dynamics algorithm, whose main idea has been put forward in \cite{Metz2016}. This numerical approach consists in representing each density $w_\rho(\Delta)$, $w_k(\Gamma)$, and $w_{\sigma}(\sigma)$ by a large collection or population containing $\mathcal{L}$ random variables, which are updated according to the algorithm explained below. After a sufficient number of updating steps, the empirical distribution of each population converges to a fixed-point distribution that solves its corresponding self-consistent equation. The calculation of averages involving $w_\rho(\Delta)$, $w_k(\Gamma)$, and $w_{\sigma}(\sigma)$ is performed by taking the arithmetic mean with the corresponding population of random variables.

 Thus, by choosing a large value of $\mathcal{L}$, the unknown distributions of eqs.  (\ref{1}-\ref{3}) are parametrized as follows:
 \begin{eqnarray}
w_{\rho}(\Delta)\leftrightarrow\{\Delta^{(\alpha)}\}_{\alpha=1}^\mathcal{L} \nonumber \\
w_k(\Gamma)\leftrightarrow\{\Gamma^{(\alpha)}\}_{\alpha=1}^\mathcal{L} \nonumber \\
w_{\sigma}(\sigma)\leftrightarrow\{\sigma^{(\alpha)} \}_{\alpha=1}^\mathcal{L}. \nonumber
   \end{eqnarray}
 Once one has set up initial values for the three populations $\{\Delta^{(\alpha)},\Gamma^{(\alpha)},\sigma^{(\alpha)} \}_{\alpha=1}^\mathcal{L}$, one performs the following steps at a single iteration of the algorithm:
\begin{enumerate}
\item  Estimate the constant $\mathcal{A}$, defined in eq. (\ref{eq:A}), as follows
\beeq{
  \frac{\alpha d}{\mathcal{A}} \simeq \frac{1}{\mathcal{L}}\sum_{\alpha=1}^{\mathcal{L}} \exp{\left[-\frac{i y}{2\pi}\text{Log}\left(\frac{1
        +\sigma^{(\alpha)} }{1+\overline{\sigma^{(\alpha)} }}\right)\right]}. \nonumber
}
\item Draw a  random number $\ell\sim\text{Poisson}(\mathcal{A})$  and select the variables $\{\Gamma^{(\alpha_s)}\}_{s=1}^\ell$ uniformly and randomly from the population $\{\Gamma^{(\alpha)}\}_{\alpha=1}^\mathcal{L}$.
\item Using the $\ell$ variables chosen in the previous step, calculate the following quantities
\beeq{
  I&=\left\lfloor e^{\frac{i y}{2\pi} W_{\epsilon}(\Gamma_1,\dots,\Gamma_\ell)}\right\rfloor,\\
  R&= e^{\frac{i y}{2\pi} W_{\epsilon}(\Gamma_1,\dots,\Gamma_\ell)}-I,
}
where $\left\lfloor \dots \right\rfloor$ is the floor function and $W_{\epsilon}$ is defined in eq. (\ref{wei}).
\item Pick up an element $\Delta^{(\alpha_{0})}$ uniformly and randomly from  the population $\{\Delta^{(\alpha)}\}_{\alpha=1}^{\mathcal{L}}$ and
  update its value as
\beeq{
\Delta^{(\alpha_0)}\to \Delta^{(\alpha_0)}=\frac{1}{\sum_{s=1}^\ell\Gamma^{(\alpha_s)}-x_\epsilon}, \nonumber
}
with rate $R$. Then increase the population to size  $\mathcal{L}+I$ by adding $I$ extra values of $\Delta^{(\alpha_0)}$ to the original population. Finally, filter the new population back to its original size  by choosing randomly $\mathcal{L}$ elements from the $\mathcal{L}+I$ available.

\item Draw a  random number $k\sim\text{Poisson}(d)$  and select $\{\Delta^{(\alpha_{\ell})}\}_{\ell=1}^k$ uniformly and randomly from the population $\{\Delta^{(\alpha)}\}_{\alpha=1}^\mathcal{L}$. Besides that, draw independently $k$ random numbers $\{ \xi_\ell \}_{\ell=1}^k$ from the distribution $P_\xi(\xi)$.
\item Pick up an element $\sigma^{(\alpha_0)}$ uniformly and randomly from $\{\sigma^{(\alpha)} \}_{\alpha=1}^{\mathcal{L}}$ and update its value according to
\beeq{
  \sigma^{(\alpha_0)} \to\sigma^{(\alpha_0)} =\frac{1}{d}\sum_{\ell=1}^k\xi^2_\ell \Delta^{(\alpha_\ell)}.
  \nonumber
}
\item Draw a single random number $\xi$ from $P_\xi(\xi)$  and choose a pair of elements $\sigma^{(\alpha_1)}$ and $\Gamma^{(\alpha_0)}$ uniformly and randomly from $\{\sigma^{(\alpha)}\}_{\alpha=1}^{\mathcal{L}}$  and $\{\Gamma^{(\alpha)}\}_{\alpha=1}^{\mathcal{L}}$, respectively. Update  $\Gamma^{(\alpha_0)}$ as follows
\beeq{
  \Gamma^{(\alpha_{0})} \to \Gamma^{(\alpha_{0})}=\frac{\xi^2}{d \left( 1+\sigma^{(\alpha_1)} \right)}.
  \nonumber
}
\item Go back to step 1 and repeat the steps 1-7 until the empirical distributions of $\{\Delta^{(\alpha)}\}_{\alpha=1}^\mathcal{L}$,  $\{\Gamma^{(\alpha)}\}_{\alpha=1}^\mathcal{L}$ and $\{\sigma^{(\alpha)} \}_{\alpha=1}^{\mathcal{L}}$ attain stationary profiles. We use the standard convention that a single Monte Carlo step consists in repeating  $\mathcal{L}$ times the steps 1-7.
\end{enumerate}
For each choice of parameters $\alpha$, $x$, $d$ and $y$, we fix $\epsilon=10^{-8}$ and $\mathcal{L}$ between $10^6$ and $10^{7}$. Such value of $\epsilon$ is sufficiently small such that the limit $\epsilon \rightarrow 0^+$ is attained. We have usually run the algorithm between $300-600$ Monte Carlo steps, which is more than sufficient to reach convergence. In some cases, in order to further improve the accuracy of our estimates, we have performed averages over independent runs of the algorithm.

One of our aims consists in deriving results for the rate function $\Psi_x(k)$. However, from the expression \eqref{eq:rf}, we are required to find the value of $y$ such that, for a fixed $k$, the following equation is fulfilled
\begin{equation}
  k=\kappa_1(y)\equiv \frac{\partial \mathcal{F}_x(y)}{\partial y}. 
  \label{ssa}
  \end{equation}
Here $\kappa_1(y)$ corresponds to the weighted first cumulant, which can be expressed  in terms of averages of the random variables $\{ \mathcal{I}_{\ell} \}_{\ell=1}^3$, defined in eqs. (\ref{11}-\ref{13}). Although it is certainly possible to find numerically $y$ obeying eq. \eqref{ssa}, it is more efficient to evaluate the rate function  $\Psi_x(k)$ parametrically in $y$: for a given value of  $y$, we determine the corresponding value of $k=\kappa_1(y)$ using eq. (\ref{ssa}). The rate function for such value of $k$  is simply given by $\Psi_x(k)= \mathcal{F}_x(y) - k y $. 

\begin{figure}[H]
\begin{center}
\includegraphics[scale=0.65]{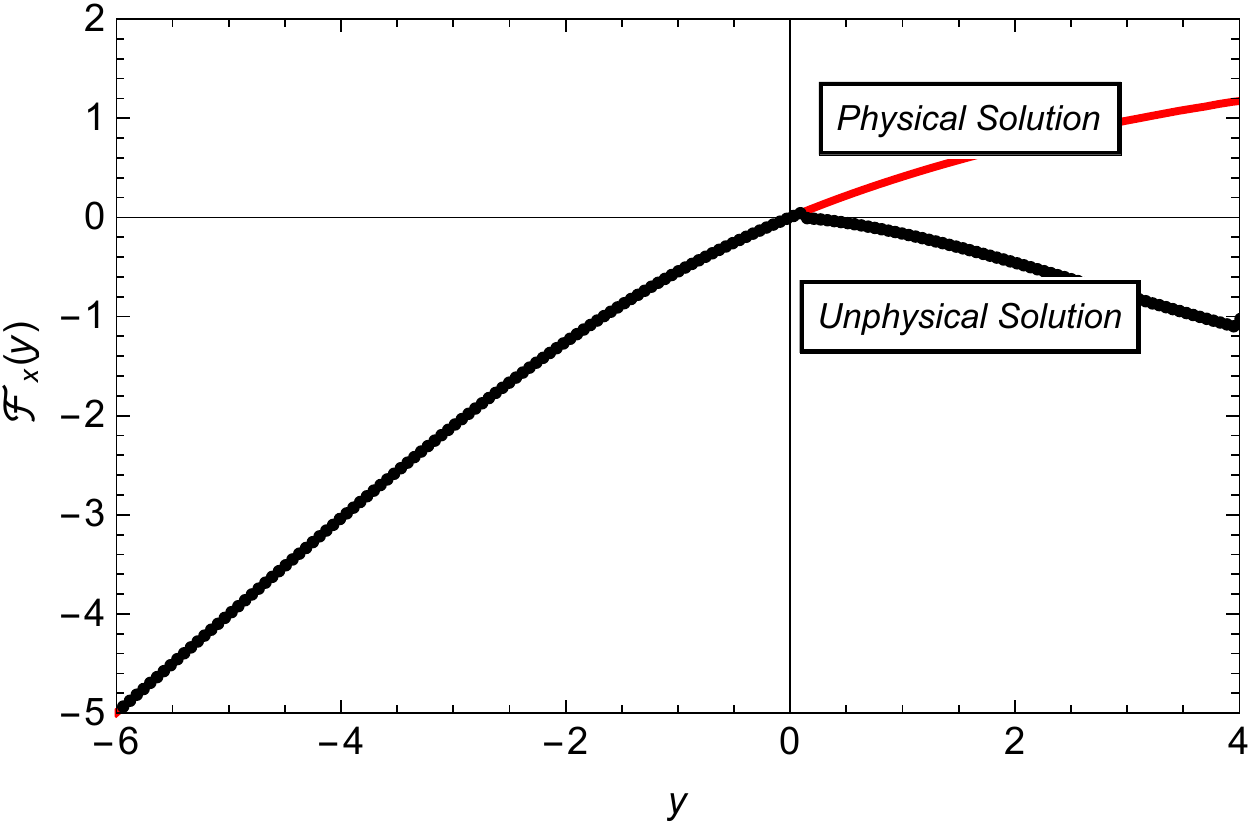}
\caption{Population dynamics results for $\mathcal{F}_x(y)$ as a function of $y$ in the case of $x= 1.01$, $\epsilon=10^{-8}$, $d=1$ and $\alpha=2$. This result has been obtained using $\mathcal{L} =10^6$ and  $300$ Monte Carlo steps. For $y>0$ there are two possible values of $\mathcal{F}_x(y)$, with the physical branch corresponding to the red curve. }
\label{fig2SI}
\end{center}
\end{figure}

When applying the weighted population dynamics algorithm, we noted that, for  positive  values of $y$, the cumulant generating function $\mathcal{F}_x(y)$ has two extrema as a functional of the distributions, that is, we find two distinct fixed-point solutions of eqs. (\ref{1}-\ref{3}) depending on the choice of the initial distributions. We illustrate this feature in figure \ref{fig2SI}, where we show both the physical and the unphysical behavior of  $\mathcal{F}_x(y)$ as a function of $y$. The physical branch is obtained by choosing the fixed-point solution  of eqs. (\ref{1}-\ref{3}) that yields $\frac{\partial \mathcal{F}_x(y)}{\partial y} \geq 0$. We have found that the  initial conditions to obtain the physical solution are such that  $\text{Im}(\Gamma)=\text{Im}(\Delta)=\text{Im}(\sigma)=0$.

\begin{figure}[H]
\begin{center}
\includegraphics[scale=0.65]{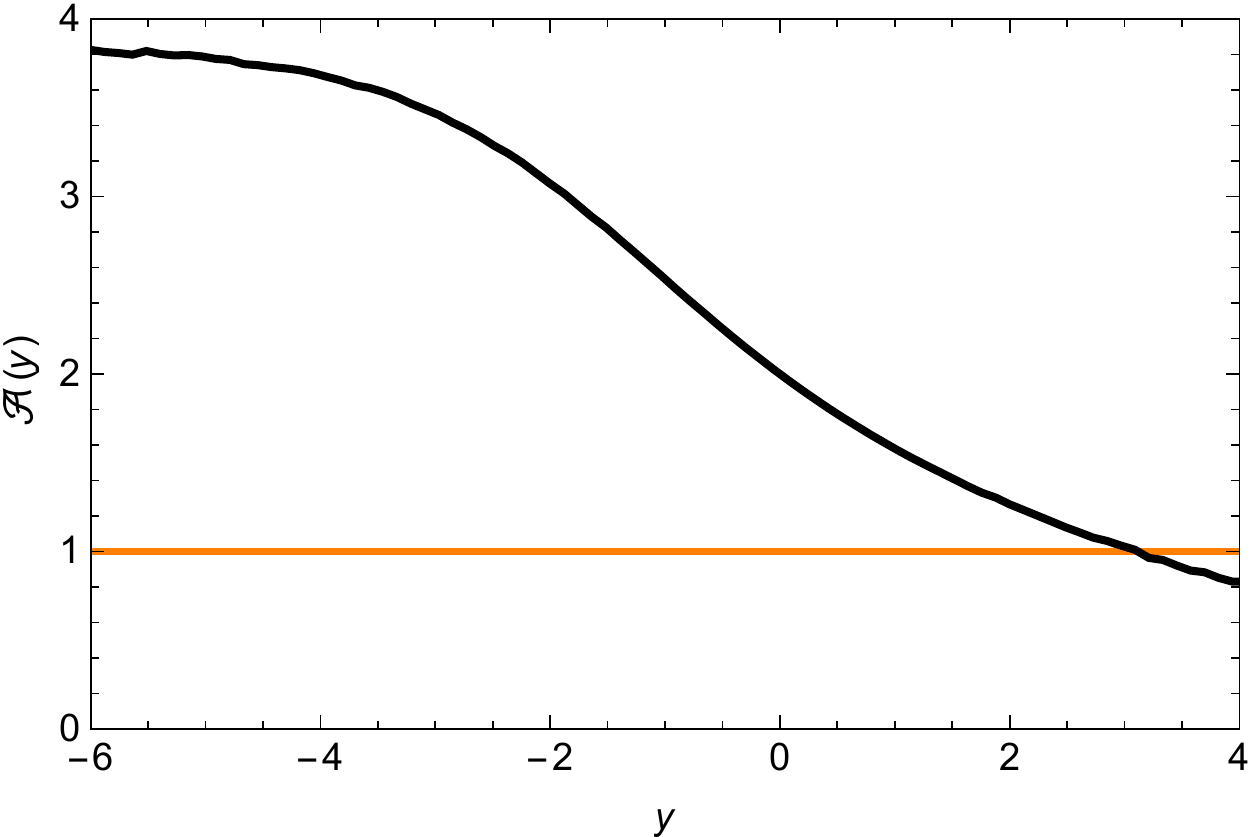}
\caption{Population dynamics results for $\mathcal{A}(y)$, defined by eq. \eqref{eq:A}, as a
  function of $y$ in the case of $x=1.01$, $\epsilon=10^{-8}$, $d=1$ and $\alpha=2$.
This result has been obtained using $\mathcal{L} =10^6$ and  $300$ Monte Carlo steps.} 

\label{fig3SI}
\end{center}
\end{figure}

To conclude this appendix, let us briefly comment on a limitation of our numerical method. The factor $\mathcal{A}$, given by eq. \eqref{eq:A}, plays the role of a rescaled  average degree $\alpha d$ and it is a function of $y$. In Figure \ref{fig3SI} we present population dynamics results for $\mathcal{A}$ as a function of $y$ and, as we can see, there is a range of $y$ for which $\mathcal{A}$ is smaller than the percolation threshold. In this situation, we have found that population dynamics does not provide reliable numerical estimates for the densities, and we had to disregard the corresponding results.

\begin{acknowledgments}
FLM and IPC thank London Mathematical Laboratory for financial support. FLM acknowledges financial support from CNPq (Edital Universal 406116/2016-4). This work has been funded by the program UNAM-DGAPA-PAPIIT IA101815. The authors thank the support of DGTIC for the use of the HP Cluster Platform 3000SL, codename Miztli,   under the Mega-project  LANCAD-UNAM-DGTIC-333.
\end{acknowledgments}

\bibliography{biblio.bib}

\end{document}